\documentclass[a4paper]{jpconf}
\usepackage{graphicx}

\newcommand{\be}{\begin{equation}}
\newcommand{\ee}{\end{equation}}
\newcommand{\bea}{\begin{eqnarray}}
\newcommand{\eea}{\end{eqnarray}}
\newcommand{\beaa}{\begin{eqnarray}}
\newcommand{\eeaa}{\end{eqnarray}}
\newcommand{\ba}{\begin{array}}
\newcommand{\ea}{\end{array}}
\newcommand{\bit}{\begin{itemize}}
\newcommand{\eit}{\end{itemize}}
\newcommand{\ben}{\begin{enumerate}}
\newcommand{\een}{\end{enumerate}}
\begin{document}
\title{DNA waves and water}

\author{L. Montagnier${}^{1,2}$, J. Aissa${}^{2}$,  E. Del Giudice${}^{3}$, C. Lavallee${}^{2}$, A. Tedeschi${}^{4}$, and G. Vitiello${}^{5}$}

\address{${}^{1}$ World Foundation for AIDS research and Prevention (UNESCO), Paris, France}
\address{${}^{2}$ Nanetics Biotecnologies, S.A. 98 rue Albert Calmette, F78350 Jouy-en-Josas, France}
\address{${}^{3}$ IIB, International Institute for Biophotonics, Neuss, Germany}
\address{${}^{4}$ WHITE HB, Milano, Italy}
\address{${}^{5}$ Dipartimento di Matematica e Informatica, Universit\`a di Salerno and\\ INFN, Gruppo Collegato Salerno,  I-84100 Salerno, Italy\\
}

\ead{vitiello@sa.infn.it}

\begin{abstract}
Some bacterial and viral DNA sequences have been found to induce low frequency electromagnetic waves in high aqueous dilutions. This phenomenon appears to be triggered by the ambient electromagnetic background of very low frequency. We discuss this phenomenon in the framework of quantum field theory. A scheme able to account for the observations is proposed. The reported phenomenon could allow to develop highly sensitive detection systems for chronic bacterial and viral infections.
\end{abstract}

\section{Introduction}
\vspace{0,5mm}

Over the last $60$ years, the development of basic knowledge in biology as well as many medical applications owes much to the discoveries made in DNA. Here is a partial list emphasizing the main advances in DNA discovery:

1944 Transformation of bacteria by DNA (O. Avery, C. McLeod, and M. McCarty)
\vspace{0,5mm}

1953 Double helix structure elucidated (J. Watson, F. Crick, M. Wilkins, R. Franklin)
\vspace{0,5mm}

1956 DNA polymerase (A. Kornberg)
\vspace{0,5mm}

1968 Restriction enzymes (W. Arber)
\vspace{0,5mm}

1969 Reverse transcription of retroviruses (H. Temin, D. Baltimore)
\vspace{0,5mm}

1976 DNA sequencing (A. Maxam, W. Gilbert, F. Sanger)
\vspace{0,5mm}

1986-1988 Polymerase chain reaction (K. Mullis)
Taq polymerase (R.K. Saiki)
\vspace{0,5mm}

2001 First Human Genome Sequence
\vspace{0,5mm}

2004-2010 High-Throughput DNA Sequencing.
\vspace{1mm}

On the other hand, in the same times evidence has been accumulated on the influence of electromagnetic  (em) fields on living organisms. The frequencies of the involved em fields cover different intervals corresponding to the different scales present in the organisms. In the present paper, by referring to recently published experimental results \cite{1,2,3},  we discuss the appearance of a new property of DNA correlated with the induction of extremely low frequency (ELF) em fields. These fields can be induced by suitable procedures in water dilutions which become able to propagate the information contained in the DNA of the original organisms to other ones.

The paper includes three Sections: the new facts, a theoretical scheme where to discuss them and the medical applications.

\section{The new facts: a new property of DNA and the induction of electromagnetic waves in water dilutions}
\vspace{0,5mm}

The story started ten years ago when one of us (L.M.) studied the strange behaviour of a small bacterium, a frequent companion of HIV, {\it Mycoplasma pirum}, and like HIV a lover of human lymphocytes. L.M. was trying to separate the bacterium, which is about $300 ~nm$ in size, from viral particles whose size is about $120 ~nm$ by filtration using filters of $100 ~nm$ and $20 ~nm$.
Starting with pure cultures of the bacterium on lymphocytes, the filtrates were indeed sterile for the bacterium when cultured on a rich cellular medium, SP4. Polymerase chain reaction (PCR) and nested PCR, based on primers derived from a gene of {\it M. pirum} which had been previously cloned and sequenced, adhesin, were negative in the filtrate. However, when the filtrate was incubated with human lymphocytes, (previously controlled for not being infected with the mycoplasma) the mycoplasma with all its characteristics was regularly recovered! Then the question was raised: what kind of information was transmitted in the aqueous filtrate?
It was the beginning of a long lasting investigation bearing on the physical properties of DNA in water.
Indeed, a new property of {\it M. pirum} DNA was found: the emission of low frequency waves in some water dilutions of the filtrate, soon extended to other bacterial and viral DNAs.

The  apparatus used to detect the electromagnetic signals comprises a solenoid capturing the magnetic component of the waves produced by the DNA solution in a plastic tube converting the signals into electric current. This current is then amplified and finally analyzed in a laptop computer using specific software (Fig.$~1$).

\begin{figure}
\begin{center}
\includegraphics{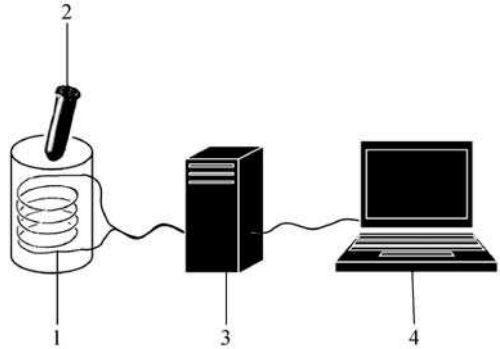}
\end{center}
\caption{\label{label} Device for the capture and analysis of em signals. (1) Coil made up of copper wire, impedance 300 Ohms. (2) Plastic stoppered tube containing 1 ml of the solution to be analyzed. (3) Amplifier. (4) Computer. From Ref. 1}
\end{figure}

\begin{figure}
\begin{center}
\includegraphics{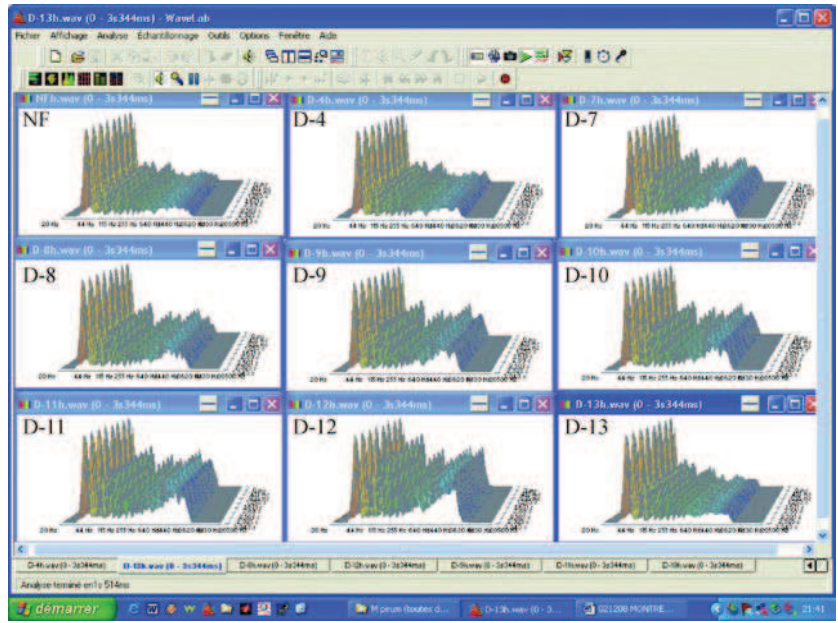}
\end{center}
\caption{\label{label} Typical signals from aqueous dilutions of {\it M. pirum} (Matlab software). Note the positive signals from
D-7 to D-12 dilutions. From Ref. 1}
\end{figure}

Here is a brief summary of the laboratory observations, which are described in more detail in \cite{1,2,3}:

1) Ultra Low Frequency Electromagnetic Waves (ULF $500-3000 ~Hz$) were detected in certain dilutions of filtrates ($100~nm$, $20 ~nm$) from cultures of micro-organisms (virus, bacteria) or from the plasma of humans infected with the same agents (Fig.$~2$). Same results are obtained from their extracted DNA.

2) The electromagnetic signals (EMS) are not linearly correlated with the initial number of bacterial cells before their filtration. In one experiment the EMS were similar in a suspension of {\it E. coli} cells varying from $10^{9}$ down to $10$. It is an all or none phenomenon.

3) EMS are observed only  in some high water dilutions of the filtrates. For example, from $10^{-9}$ to $10^{-18}$ dilutions in some preparations of {\it E. coli} filtrates.

4) In the case of {\it M. pirum}, an isolated single gene (adhesin, previously cloned and sequenced) could induce the EMS. As the gene was cloned in two fragments, each of the isolated fragments was able to generate EMS, suggesting that a short DNA sequence was sufficient to induce the signals. Similarly,  a short HIV DNA sequence ($104$ base pair) is found to be  sufficient to produce the EMS.

5) Some bacteria are not producing EMS: this is the case of probiotic bacteria such as {\it Lactobacillus} and also of some laboratory strains of {\it E. coli} used as cloning vector.

6) These studies have been extended to viruses, although not all the viral families have been explored. Similar EMS were detected from some exogenous retroviruses (HIV, FeLV), hepatitis viruses (HBV, HCV), and influenza A (in vitro cultures). In general, EMS are produced by $20 ~nm$ filtrates of viral suspensions or from the extracted DNA: a question remains for RNA viruses (HCV, influenza) as to whether the RNA from the mature viral particles is a source of EMS, or not. In the case of HIV, EMS are not produced by the RNA of viral particles, but rather are produced by the proviral DNA present in infected cells.
In the case of bacteria, EMS are produced by $100 ~nm$ filtrates and not by $20 ~nm$ filtrates, indicating that the size of the structures producing EMS is ranging between $20$ and $100 ~nm$. This justifies the name of nanostructures.
These studies are highly suggestive that one is dealing with nanostructures made of water. Highly purified water samples are used, although one cannot exclude the role of minimal traces of impurities.
The EMS production by the nanostructures is resistant to: Rnase treatment, Dnase (while this will destroy the DNA at the origin of EMS), Protease (proteinase K), Detergent (SDS). However, they are sensitive to heat (over $70 ~{}^{\circ}C$) and freezing ($-80 ~{}^{\circ}C$). This sensitivity is reduced when dealing with purified short DNA sequences.
The technical conditions for EMS induction is summarized by the following list:

- Filtration: $\frac{450}{100} ~nm$ for bacterial DNA, $\frac{450}{20} ~nm$ for viral DNA

-  High dilutions in water

-  Mechanical agitation (Vortex) between each dilution

- Excitation by the electromagnetic background of extremely low frequency (ELF), starting very low at $7 ~Hz$. The excitation is not induced when the system is shielded by a  mu-metal cage.

The stimulation by the electromagnetic background of very low frequency is essential. The background is either produced from natural sources (the Schumann resonances \cite{4a}
which start at $7.83 ~Hz$) or from artificial sources.

\begin{figure}
\begin{center}
\includegraphics{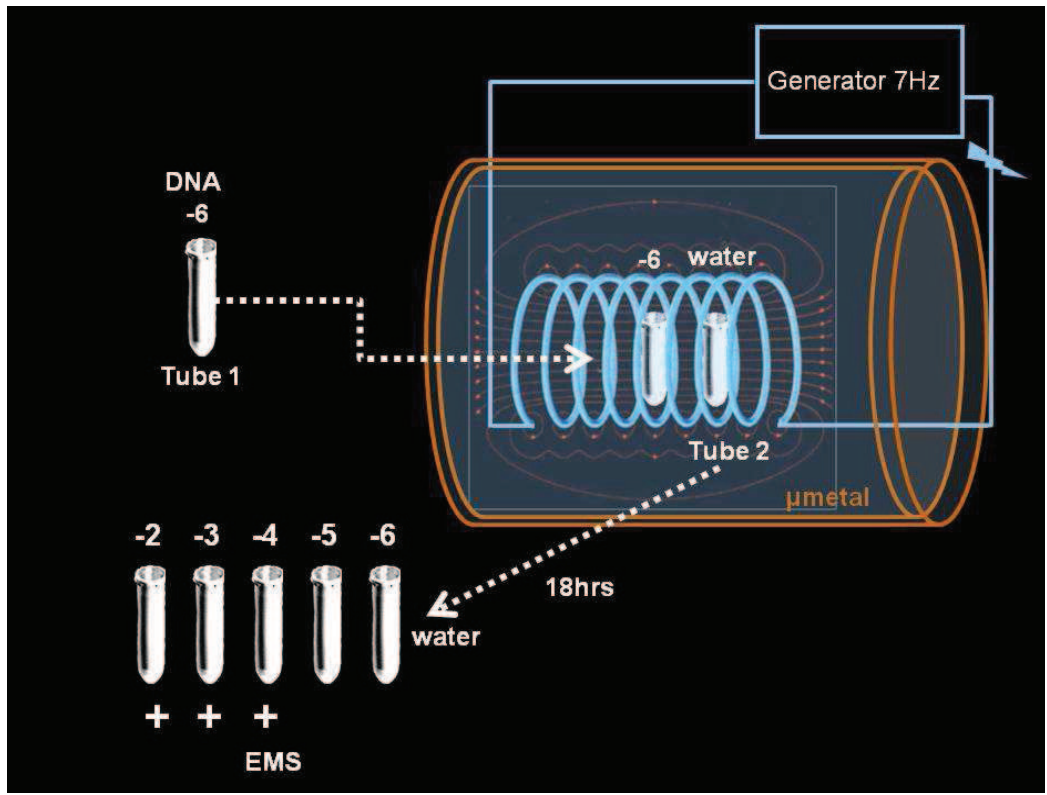}
\end{center}
\caption{\label{label}Transmission of DNA genetic information into water through electromagnetic waves. From Ref. 3}
\end{figure}


\subsection{Transmission of DNA sequence through waves and water}
\vspace{0,5mm}

In further experiments  a fragment of HIV DNA taken from its long terminal repeat (LTR) has been used as the DNA source. This fragment was amplified by PCR ($487$ base pairs) and nested PCR ($104$ base pairs) using specific primers. In a first step, DNA dilutions were made, in which the production of EMS under the ambient electromagnetic background was detected. Then the following steps were taken.
As shown in Fig.$~3$, one of the positive dilutions (say $10^{-6}$) was placed in a container shielded by $1 ~mm$ thick layer of mu-metal (an alloy absorbing ultralow frequency waves). In its vicinity another tube containing pure water was placed. The water content of each tube was filtered through $450 ~nm$ and $20 ~nm$ filters and diluted from $10^{-2}$ to $10^{-15}$. A copper solenoid is placed around them and receives a low intensity electric current oscillating at $7 ~Hz$, produced by an external generator. The produced  magnetic field is maintained for $18$ hours at room temperature. EMS are then recorded from each tube. Now also the tube containing water emits EMS, at the dilutions corresponding to those positive for EMS in the original DNA tube.
This result shows that, upon $7 ~Hz$ excitation, the transmission into pure water of the oscillation of the nanostructures initially originated from DNA has been achieved.
The following controls were found to suppress the EMS transmission in the water tube:

- Time of exposure of the two tubes less than $16-18$ hrs
\vspace{0,5mm}

- No coil
\vspace{0,5mm}

- Generator of magnetic field turned off
\vspace{0,5mm}

- Frequency of excitation $< ~7 ~Hz$
\vspace{0,5mm}

- Absence of DNA in tube $1$.
\vspace{0,5mm}

At this point the most critical step was undertaken, namely to investigate the specificity of the induced water nanostructures by recreating from them the DNA sequence.
For this all the ingredients to synthesize the DNA by polymerase chain reaction (nucleotides, primers, polymerase) were added to the tube of signalized water. The amplification was performed under classical conditions ($35$ cycles) in a thermocycler. The DNA produced was then submitted to electrophoresis in an agarose gel. The result was that a DNA band of the expected size of the original LTR fragment was detected.
It was further verified that this DNA had a sequence identical or close to identical to the original DNA sequence of the LTR. In fact, it was $98 ~\%$ identical ($2$ nucleotide difference) out of $104$.
This experiment was found to be highly reproducible ($12$ out of $12$) and was also repeated with another DNA sequence from a bacterium, {\it Borrelia burgdorferi}, the agent of Lyme disease. It was shown clearly  that the water nanostructures and their electromagnetic resonance can faithfully perpetuate DNA information.
These elements give support to a provocative explanation of our {\it Mycoplasma pirum} filtration experiment (Fig.$~1$): the nanostructures induced by {\it M. pirum} DNA in the filtered water represent different segments of its genomic DNA. Each nanostructure when in contact  with the human lymphocytes is retro-transcribed in the corresponding DNA by some cellular DNA polymerases. Then there is a certain probability (even very low) that each piece of DNA recombines within the same cell to other pieces for reconstructing the whole DNA genome.
We have to assume that in presence of the eukaryotic cells the synthesis of the mycoplasma components (membrane lipids, ribosomes) can be also instructed by the mycoplasma DNA. One single complete mycoplasma cell is then sufficient to generate the whole infection of lymphocytes. Recent experiments of the G. Vinter group have shown \cite{4b} that a synthetic genomic DNA is sufficient to maintain all the characteristics of a mycoplasma.
All the steps assumed in the regeneration from water can be analyzed and open to verification.

\section{The theoretical framework}
\vspace{0,5mm}

The above experimental observations fit into the  physical view which addresses biological dynamics as an interplay of chemical processes and em interactions; in other words, as an array of em assisted biochemical reactions.

We will try to interpret  the above experimental results in the framework of a recently proposed theory of liquid water based on Quantum Field Theory (QFT) \cite{4}-\cite{9}.
This theory is intrinsically non-linear and it provides the suitable tools to describe a complex ensemble of processes which are also non-linear.

Let us first summarize the main points of such a theory, whose details can be found in the quoted references.

The starting point is the realization that liquid water molecules cannot be assumed to be bound by purely static interactions (H-bonds, electric dipole-dipole interaction).  Their binding is actually induced by the time-dependent radiative long range em field. Short range static bonds, such as $H$-bonds, then set in as a consequence of the molecule condensation induced by such long range radiative fields.

The main results of the theory  are \cite{4}-\cite{9}:
\vspace{1mm}

$a$) An ensemble of molecules interacting with the radiative em field acquires, above a density threshold and below a critical temperature,  a new non-trivial minimum energy state, different from the usual one where the oscillations of the molecules are uncorrelated and the em field is  vanishing. The new minimum energy state implies a configuration of the system where all molecules enclosed within an extended region, denominated Coherence Domain (CD), oscillate in unison in tune with an em field trapped within the CD. The size of this extended region is just the wavelength $\lambda$ of the trapped em field. The collective coherent oscillation of the molecules component the CD occurs between the individual molecule ground state and an excited state whose volume, according to atomic physics, is wider than the ground state volume. The wavelength $\lambda$ of the trapped em field depends on the excitation energy $E_{exc}$  through the equation:
\be \label{1}
\lambda \, =\, \frac{hc}{E_{exc}}
\ee

The CD is a self-produced cavity for the em field because of the well known Anderson--Higgs--Kibble mechanism \cite{7} which implies that the photon of the trapped em field acquires an imaginary mass, becoming therefore unable to leave the CD. It is just this self-trapping of the em field that guarantees that the CD energy has a finite lower bound. Because of this self-trapping the frequency of the CD em field becomes much smaller than the frequency of the free field having the same wavelength.
The above results apply to all liquids. The peculiarity of water is that the coherent oscillation occurs between the ground state and an excited state lying at $12.06 ~eV$  just below the  ionization threshold ($12.60 ~eV$). In the case of liquid water, the CD (whose size is $100 ~nm$ according to Eq.~(\ref{1})) includes an ensemble of almost free electrons which are able to accept externally supplied energy and transform it into coherent excitations (vortices) whose entropy is much lower than the entropy of the incoming energy. Consequently, water CDs could become dissipative structures in the sense of the thermodynamics of irreversible processes \cite{10}-\cite{12}.

\vspace{1mm}

$b)$	Coherence among molecules is counteracted at any non-vanishing temperature T by thermal collisions which could put molecules out of tune, like in the Landau picture of liquid Helium \cite{13}. The competition between electrodynamic attraction and thermal noise produces a permanent crossover of molecules between a coherent regime and a non-coherent one. For a given value of T, the total number of coherent and non-coherent molecules are constant, but each molecule oscillates between the two regimes producing a continuous change of the space distribution of the coherent and non-coherent fractions of the molecules. It is just this flickering landscape of the two phases of liquid water  to produce, in the case of experiments whose resolution time is long enough, the appearance of water as a homogeneous liquid. However, the above property holds only for bulk water. Near a surface, the attraction between water molecules and the surface could protect the coherent structure from the thermal noise, giving rise to a stabilization of the coherent structure. This is in particular the case of living organisms where water molecules are bound to membranes or biomolecule backbones. In this case CDs live long enough to exhibit the peculiar properties of coherence.

\vspace{1mm}

$c)$	CDs store externally supplied energy in form of coherent vortices. These vortices are long lasting because of coherence, so that a permanent inflow of energy produces a pile up of vortices; they sum up to give rise to a unique vortex whose energy is the sum of the partial energies of the excitations which have been summed up. In this way water CDs can store a sizeable amount of energy in a unique coherent excitation able to activate molecular electron degrees of freedom; this high-grade energy is the sum of many small contributions, whose initial entropy was high.

\vspace{1mm}

$d)$ CDs oscillate on a frequency common to the em field and the water molecules  and  this frequency changes when energy is stored in the CD. When the oscillation frequency of the CD matches the oscillation frequency of some non aqueous molecular species present on the CD boundaries, these ``guest" molecules become members of the CD and are able to catch the whole stored energy, which becomes activation energy of the guest molecules; consequently, the CD gets discharged and a new cycle of oscillation could start. The above mechanism fits the intuition of Albert Szent-Gyorgyi \cite{14} who proposed half a century ago that water surrounding biomolecules should be at the origin of the excitations of molecule electron levels responsible for chemical reactions.
Moreover, should the ensemble of frequencies able to attract the component monomers of a polymer be excited in the water CD, the polymer would be created by the attraction of the monomers on the CD, provided that they be present in the solution. In this way it is possible to induce the polymerization of monomers by supplying to the water CDs of the monomer solution the  em fields having the relevant frequencies (electromagnetic information).

\vspace{1mm}

$e)$ A collective performance of water CDs, which could give rise to a biochemical activity synchronous in a mesoscopic region, should demand a uniform rate of energy loading for all the involved CDs. This requirement is satisfied by a mechanism which includes electrolyte ions, whose essential role in the biological dynamics is widely recognized. Ions close to water CDs are attracted by the em field trapped in the domains; so they are kept orbiting around the domain moving at a circular speed proportional to the so called cyclotron frequency  $\nu_c$ :
\be \label{2}
\nu_c \,= \,\frac{1}{2\pi}\frac{q}{m} B
\ee
where $q$ and $m$ are the electric charge and the mass of the ion, respectively, and $B$ is the magnetic field. Since DNA and also proteins are polyelectrolytes, they are surrounded by a cloud of positive counter-ions; ions having a cyclotron frequency in the interval between 1 and $100 ~Hz$ play an important role. It has been experimentally detected by M. Zhadin \cite{15} and later by Zhadin and Giuliani \cite{16}  that by applying a magnetic field, having a frequency which  matches the ion cyclotron frequency, on a system where ions are present, these ions are extracted from their orbits. This mechanism has been theoretically clarified in \cite{17}. Due to the conservation of angular momentum, the extraction of ions from the cyclotron orbits produces a rotational motion of the quasi free electrons of the water CDs, which therefore become energetically excited \cite{18}. In the case where the ion concentration could be assumed to be uniform in a mesoscopic region and the externally applied magnetic field has also a mesoscopic size, the amount of energy excitation could be assumed uniform in a region including a large number of water CDs which correspondingly are excited in a uniform way, thus ensuring the coherence among them. The persistence in time of such extremely low frequency magnetic fields guarantees a steady excitation of the water CDs and correspondingly of the biochemical activity catalyzed by them.

\vspace{1mm}

Let us analyze now the experimental results reported in Section $2$ within the theoretical framework summarized above.

The role played by the background of low em frequency is understood by observing that
in order to load energy in the water CDs, we need a resonant alternating magnetic field. In higher organisms, such as the humans, this field can be produced by the nervous system. Elementary organisms, such as bacteria, should use environmental fields. Good candidates are the Schumann modes of the geomagnetic
field \cite{4a}. These modes are the stationary modes produced by the magnetic activity (lightnings or else) occurring in the shell whose boundaries are  the surface of the Earth and the conductive ionosphere, which acts as a mirror wall for the wavelengths higher than several hundreds of meters. These stationary modes should have a frequency $\nu_{s}$ which in the ideal case is \cite{4a}
\be \label{Schum}
\nu_{s} (n) \,= \,\frac{c}{2\pi R} \sqrt{n(n+1)}~,
\ee
where $R$ is the radius of the Earth. The real Earth-ionosphere cavity is not an ideal one, so that real frequencies are a little bit lower than the values given by Eq.~(\ref{Schum}). Actually, peaks around $7.83$, $14.3$, $20.8$, $27.3$ and $33.8$ $Hz$ are experimentally found.

Consequently, in order to produce the energy loading of CDs, the biological system should select ion species having a $\frac{q}{m}$ ratio such that, given the local value of $B$ in the organism, the value given by Eq.~(\ref{2}) fits with one of the Schumann resonances. The local value of $B$ is expected to be not much different from the value of the Earth's magnetic field, which is in the order of $50$ microtesla.

As explained in the point $e$ above, ions are extracted from their orbits  when a magnetic field $B$ is applied so to match the ion cyclotron frequency. This is done indeed by the Schumann mode of $7.83 ~Hz$ of the geomagnetic field.
Such an  extraction of ions from their cyclotron orbits produces in turn,  because of conservation of angular momentum, a counter-rotation of the plasma of quasi-free electrons in the CDs, whose frequency depends, of course, on the number of involved ions, namely on their concentration, which therefore is the only relevant variable. The phenomenon occurs in the same way on all the CDs of the system, whose number is irrelevant for this purpose, in agreement with the item $2$ in Section $2$.

As a result, the so induced rotation of the  plasma of the quasi-free electrons in the CDs produces the observed EMS signal.

A confirmation that the observed frequencies are produced by magnetic activity is given by the circumstance that the excitation is prevented by mu-metal absorption.

Summarizing, the above analysis fits with the role that the frequency at $7.83 ~Hz$ plays in the observations (see Section $2$).

It is interesting to observe that, should the cyclotron orbits around the water shell be saturated by an ion species which does not  match the Schumann resonances, the activity of the biological system would be inhibited. This prediction is in agreement with facts since we know that there are ions promoting biological activity and ions inhibiting it. The above conclusion holds, of course, if the only em background is the natural one (Schumann modes) or an artificial em background with frequencies similar to the Schumann ones; should an artificial em background with a different frequency distribution be present, a reshuffling of the favorable and unfavorable ion species would occur. This feature could provide a rationale for the observed impact of ELF fields on the physiological activity.

We  observe also that the dependence on the concentration is at the origin of the dependence of the signal frequencies on the aqueous dilution. Since angular momentum is quantized, a threshold in the dilution (ion concentration) value is naturally expected, as indeed observed.
In more detail, suppose one is able to extract $n$ ions per CD, then the plasma of $N$ quasi-free electrons starts to rotate with a frequency much higher than the ion cyclotron frequency since electron mass is much smaller than the ion mass. A corresponding co-resonating field appears in the surroundings of the rotating CDs and could be at the origin of an extended coherence among CDs.

The CD frequency is decreased by the increase of aqueous dilution. The existence of a window of dilutions (see point $3$ in Section $2$) could be understood by presuming that the signal produced by the lower dilutions could have a frequency higher than the interval of the values detectable by the used instruments. Higher dilutions, on the contrary, could  produce no signal because the ion concentration is decreased below the threshold able to excite the CDs.

We note that the coherence among the CDs in a region of mesoscopic extension, produced in the presence of the excitation energy in such a region, may  sustain the non-dispersive propagation of the em radiative field.
Here a peculiar role is played by the specific DNA structure and by the coherent water surrounding it. This is a point to be further analyzed on the basis of the specific chemical structure of the DNA fragments used in the experiments. Such an important investigation is in our future plans.

The field associated to the observed EMS, generated through the highly non-linear mechanism described above,
may also trigger the coherence among the CDs in a second tube containing pure water where it is allowed to penetrate under the conditions set in the experiments described in subsection $2.1$. This phenomenon is similar to the  proximity effect observed in two superconducting samples or in the arrays of Josephson junctions, by which the samples or the junctions fall into a phase-locking regime. In the case under study, such a phase-locking regime manifests itself in the observed transmission of the excited em field of the water microstructures  which surrounded the DNA in the first tube. Once the water CDs of the second tube are excited by the em field coming from the first tube, the DNA is fabricated according to the process suggested in the point $d$ of the present section.
The phase-locking at the specific frequency of the em field propagating in the original DNA tube is clearly reflected in the specificity of the induced water microstructures out of which the original DNA sequence ($98~\%$ identical) is recreated. On the other hand, the observed high reproducibility of this experiment finds its explanation in the high stability of the coherent structures (CDs and coherent clusters of CDs) into play.
A series of experiments aimed at verifying the correctness of the described theoretical scheme have recently been undertaken.

Finally, we observe that, at the present stage of the theoretical analysis, our discussion can only lead us to qualitative agreement with the features observed in the experiments. For a quantitative fitting we need to introduce specific models in the frame of the general theoretical scheme depicted above. We leave this to a future work.

\section{Conclusions and medical applications}
\vspace{0,5mm}

In this paper we have described the experiments showing a new property of DNA and the induction of electromagnetic waves in water dilutions. We have briefly depicted the theoretical scheme which can explain qualitatively  the features observed in these experiments.

We remark that it is possible to detect the same EMS from the plasma of patients suffering from various infections and chronic diseases. The plasma has to be kept fresh and unfrozen. If the plasma is frozen at $ - 70 ~{}^\circ C$ then one must extract the DNA in order to recover the signals. The DNA can be also extracted from tissue biopsies. The list of diseases in which EMS have been found (such as Alzheimer,  Parkinson, Multiple Sclerosis, various neuropathies, chronic Lyme syndrome, rheumatoid arthritis) indicate clearly that their presence is not limited to diseases known to be of infectious origin: The fact that EMS have been found in diseases not known to be of infectious origin
is intriguing, and leads us to seek bacterial or viral factors in these diseases.
A special case is that of HIV. Signals have been regularly detected coming from HIV DNA sequences in the blood of patients treated by antiretroviral therapy and responding well to that treatment by the disappearance of viral RNA copies in the circulating blood. This would indicate that such DNA comes from a reservoir not accessible to the classical treatment, and not from viral particles circulating in the blood. Moreover, not only the plasma of the patients, but also the red blood cell fraction contains DNA emitting signals. This is intriguing, as the red cells do not contain any cellular DNA, and as the virus does not bind to erythrocyte membrane. The possibility that a third element is involved is under investigation. The hypothesis \cite{2} has been put forward that the antiretroviral treatment, which includes reverse transcriptase inhibitors, is itself selecting for a new way of viral DNA replication involving one or several cellular DNA polymerases. As for the {\it M. pirum} DNA, it is suggested that the HIV DNA fragments and their nanostructures present in the blood may not originate from cell lysis but, on the contrary, represent pieces of definite size able to recombine in the appropriate recipient cells (lymphocytes) to form whole genomic DNA and finally regenerate infectious virus.
Whatever is the origin of this DNA, its easy detection by electromagnetic signals may render it a unique biomarker for attacking the viral reservoir. Its existence in the blood by more classical PCR technology has also been confirmed \cite{2}.
One thus has a powerful tool to test new types of treatment aimed at eradicating HIV infection, which has never yet been achieved, by the treatments actually in use. This is particularly important for patients in countries in which HIV prevalence is very high, ($5 - 10 ~\%$ in a large part of sub-Saharan Africa).
The process of launching clinical trials in West and South Africa to test new therapeutics is planned. Their efficacy will be monitored by this new test, together with the improvement of more classical parameters, evaluating the full restoration of the immune system. The objective here is eradication of HIV infection, so that it will not be necessary for patients to be treated for life by a combination of toxic and expensive drugs.
Our work is interdisciplinary, involving biologists, physicists, and medical doctors. There are of course many unresolved questions raised by our findings, which deserve more work and more interactions. DNA signalling is stimulated by $7 ~Hz$ naturally occurring waves on earth. Waves produced by the human brain are also in the range of $7 ~Hz$. These are interesting questions to be asked and possibly answered.

\section*{References}

\end{document}